%% file: main.tex
\documentclass[10pt,letterpaper,compsoc,conference]{iiswc25}
 
\usepackage{cite}
\usepackage{amsmath,amssymb,amsfonts}
\usepackage{algorithmic}
\usepackage{graphicx}
\usepackage[dvipsnames]{xcolor}
\usepackage[final]{microtype}
\usepackage[italic]{mathastext}
\usepackage{libertine}
\usepackage[T1]{fontenc}
\usepackage{textcomp}
\usepackage[varqu,varl]{zi4}
\usepackage[all]{nowidow}
\usepackage[auth-lg,affil-it]{authblk}
\usepackage[keeplastbox]{flushend}
\usepackage{fancyhdr}
\usepackage{booktabs}
\usepackage{tabularx}
\usepackage{graphicx}
\usepackage{subcaption}
\usepackage[]{hyperref}
\usepackage{comment}
\usepackage{xcolor}
\usepackage[commandnameprefix=always]{changes}

\fancypagestyle{firstpage}{
  \fancyhf{}
  
  \fancyfoot[C]{\thepage}
}

\begin{document}

\newcommand{\para}[1]{\textbf{#1:}}
\newcounter{note}[section]
\renewcommand{\thenote}{\thesection.\arabic{note}}

\newcommand{\mengyuan}[1]{\refstepcounter{note}{\bf\textcolor{olive}{$\ll$mengyuan~\thenote: {\sf #1}$\gg$}}}

\newcommand{\bheading}[1]{{\vspace{4pt}\noindent{\textbf{#1}}}}

\newenvironment{packeditemize}{
\begin{list}{$\bullet$}{
\setlength{\labelwidth}{0pt}
\setlength{\itemsep}{2pt}
\setlength{\leftmargin}{\labelwidth}
\addtolength{\leftmargin}{\labelsep}
\setlength{\parindent}{0pt}
\setlength{\listparindent}{\parindent}
\setlength{\parsep}{1pt}
\setlength{\topsep}{1pt}}}{\end{list}}

%% EDIT TITLE BELOW

\title{Characterization of GPU TEE Overheads in Distributed Data Parallel ML Training }

%% DO NOT EDIT THE FOLLOWING

\renewcommand\Authsep{\qquad}
\renewcommand\Authand{\qquad}
\renewcommand\Authands{\qquad}

%% EDIT AUTHOR LIST BELOW

\author{Jonghyun Lee}
\author{Yongqin Wang}
\author{Rachit Rajat}
\author{Mengyuan Li}
\author{Murali Annavaram}
\affil{Ming Hsieh Department of Electrical Computer Engineering, University of Southern California}
%\author{Author3 Name}
%\affil{Full Name of Awesome School}

%%% ALTERNATIVE FORMAT FOR MULTIPLE SCHOOLS:
%%% 
% \author[1]{Author1 Name}
% \author[2]{Author2 Name}
% \author[2]{Author3 Name}
% \author[1]{Author4 Name}
% \affil[1]{Full Name of Awesome School}
% \affil[2]{Full Name of Awesomer School}

\maketitle
\thispagestyle{firstpage}
\pagestyle{plain}

%% EDIT YOUR PAPER'S CONTENTS BELOW

\begin{abstract}
\input{0_abstract}
\end{abstract}

\section{Introduction}
\label{sec:intro}
\input{1_introduction}

\section{Background}
\label{sec:background}
\input{3_background}

\section{Understanding DDP's Overhead in Multi-GPU TEE Systems}
\label{sec:overview}

\input{4_overview}

\section{Experimental Design}
\label{sec:method}
\input {5_methodology}

\section{Analysis on DDP Training Experiments }
\label{sec:experiment}
\input{6_evaluation}

% related works go to the end for architecture conferences

\section{Related Works}
\label{sec:related}
\input{7_relatedwork}

\section{Conclusion}
\label{sec:conclusion}
\input{8_conclusion}
\bibliographystyle{IEEEtranS}
\bibliography{reference}

\end{document}

%% file: 0_abstract.tex
Trusted Execution Environments (TEEs) have become a leading approach for securing computation in untrusted clouds. NVIDIA’s recent GPU TEEs enable training ML models while preserving model and data confidentiality, yet their impact on distributed data-parallel (DDP) training is underexplored. We present an in-depth empirical study of DDP with GPU TEEs and expose the main bottlenecks. During back-propagation, DDP performs ring all-reduce (scatter-reduce followed by all-gather). Because only the CPU and GPU packages are trusted, every message that crosses a package boundary must be encrypted and authenticated. Consequently, each sub-step of ring all-reduce performs cryptographic work on both sender and receiver; as the number of GPUs grows, the number of sub-steps and their TEE overheads scale linearly. We further show that asynchronous gradient exchange, which normally overlaps communication with computation, yields little benefit in GPU TEEs: the cost of securing and verifying many small transmissions dominates, especially for large models such as GPT-2. On four GPUs, we measure an average 8.68× increase in per-iteration time across benchmarks and a 41.6× slowdown for GPT-2-XL versus non-TEE DDP. As a mitigation, we enlarge the DDP bucket\_cap\_mb to batch more gradients per transfer, which cuts round-trip counts; this tuning narrows but does not close the gap, with GPT-2-XL on four-GPU TEEs remaining 3.03× slower than its unsecured baseline. Our results highlight the need for TEE-aware communication designs beyond parameter tuning.

%% file: 1_introduction.tex
In recent years, machine learning providers have shifted their training workloads to cloud platforms, leveraging the scalability and speed of multi-GPU clusters to handle ever-larger models. However, entrusting this process to a cloud service provider (CSP) creates serious privacy and intellectual-property concerns: a curious or compromised CSP with hypervisor-level access can easily inspect or exfiltrate proprietary model weights and training data.

NVIDIA Confidential Computing (CC)~\cite{nvidiacc,nvidiappcie,nvidiah100,nvidiab200,nvidiartxpro6000} has emerged as a leading solution to address these trust and security concerns in cloud environments. NVIDIA CC is built upon hardware-enforced isolation technology known as a Trusted Execution Environment (TEE), which leverages the hardware System-on-Chip (SoC) as the root of trust. Specifically, modern server-grade CPUs and GPUs with TEE support can create an isolated, hardware-protected execution environment for cloud computing tasks. Such hardware-based protection ensures that even a malicious or compromised hypervisor with root privileges cannot violate the confidentiality and integrity of cloud workloads executed within these protected environments.
Recent generations of NVIDIA server-grade GPUs, including the H100 (Hopper architecture)~\cite{nvidiah100}, and the B200~\cite{nvidiab200} and RTX Pro 6000 (Blackwell architecture)~\cite{nvidiartxpro6000}, natively support NVIDIA CC, significantly expanding the availability of GPU-based TEE. Additionally, public CSPs are progressively integrating TEE-protected computing into their service offerings. Major cloud providers such as AWS, Azure, and Google Cloud now all offer confidential virtual machine (CVM) services with CPU-TEE protection. Azure has also recently begun supporting CVMs equipped with NVIDIA CC-enabled GPUs, providing customers with a comprehensive, end-to-end confidential secure ML system for GPU-intensive workloads~\cite{azurecvm}.

\begin{figure}
    \centering
    \includegraphics[width=\columnwidth]{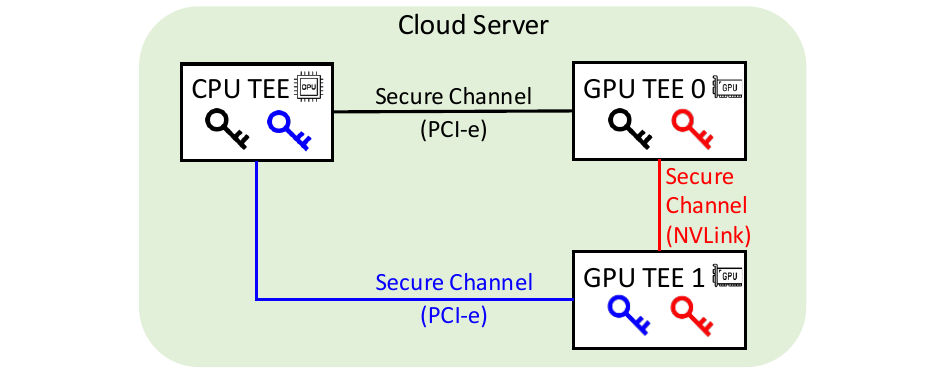}
    \caption{Multi-GPU Confidential Computing system topology. As NVLink and PCIe are not trusted, secure channel is established via shared key between two entities: CPU-to-GPU (black and blue key) and GPU-to-GPU (red key). }
    \label{fig:multigpusystemtopo}
\end{figure}

\bheading{Confidential GPU System Topology.} Figure~\ref{fig:multigpusystemtopo}  illustrates the typical system topology for confidential computing using one or multiple GPU TEEs. The topology consists of a CPU TEE responsible for network communication and GPU control, alongside GPU TEEs dedicated to secure ML tasks. 
In this design, a CPU TEE, implemented using CVM (e.g., Intel TDX~\cite{tdx} or AMD SEV-SNP~\cite{amdcc}), handles networking and control functions, and loads the GPU driver within its secure enclave. Each GPU TEE provides hardware-enforced isolation, ensuring the confidentiality and integrity of computations performed on the GPU.
Although the hardware-based isolation in CPU and GPU TEEs individually secures computations, data transfers between the CPU TEE (CVM) and GPU TEEs, or among GPU TEEs themselves, require additional encryption and data authentication to defend against bus-snooping attacks. GPUs connect through PCIe and NVIDIA’s proprietary NVLink interface, both of which are vulnerable to interception or tampering. Therefore, all transmitted data must be encrypted. This additional protection, together with TEE protections, establishes a secure multi-GPU system in the cloud. Only the CPU and GPU packaging are trusted in this system, establishing a robust threat model. 

%1. why distributed GPU CC environment is worthy doing a characterization work (because it will have more overhead).
However, establishing secure communication via encryption and authentication comes with a non-negligible price. Prior works ~\cite{wang2024fastrack, tan2024performance, tan2025pipellm} demonstrate significant performance degradation by TEE-incurred encryption overheads between CPU and GPU data transmission over PCIe. However, we must characterize the performance impact of modern ML training scenarios involving multiple GPUs, given their prevalence. In this work, we analyze the TEE overheads in GPU-to-GPU synchronization operations in multi-GPU ML training, as gradient synchronization requests occur more frequently with large data volume than infrequent CPU-GPU input transfers studied in prior works.

% 2. we did a what characterization study (emphasize the completeness and comprehensive) 
\bheading{TEE Overhead in Multi-GPU ML Training.} Among various parallelization models to enable large-scale ML training, we focus on Distributed Data Parallelism (DDP), a synchronous distributed training, which is the most widely used due to its ease of implementation and broad applicability across various benchmarks~\cite{pytorchddp}. In DDP, all participating GPUs start each training iteration with non-overlapping input batches, and synchronize their gradients during back propagation via all-reduce operations, by default ring all-reduces. We conducted a characterization study over DDP training in multiple GPU-TEEs using a wide range of ML benchmarks and a varied number of GPUs. 

%3. our results We identifies the major overhead factors a.XXXX, b. YYYY, with details in 1-2 sentences, some rational and more from an impact perspective.
Based on our work, we identify the major contributors of the multi-GPU TEE overheads as well as mitigation techniques that can be easily deployed. Firstly, ring all-reduce operations, which are used as default DDP synchronization operations between GPUs, are often broken down into multiple steps of sub-operations: scatter-reduce and all-gather. Each step of these sub-operations must encrypt and generate MAC tags on outgoing gradients, as well as decrypt and verify the MAC tags on incoming gradients. We also note that the number of scatter-reduce and all-gather operations increases linearly with the number of participating GPUs, which further impacts the scalability of GPU CC-based training. Lastly, the model size greatly determines the frequency of asynchronous gradient exchange per layer. Larger models evaluated in our work, such as GPT2, suffer from significant TEE overheads as more all-reduce operations take place. In addition to defining the cause of TEE overheads, we also suggest a lightweight technique to mitigate TEE overhead, in which we modify the DDP training hyperparameter, namely bucket\_cap\_mb, to change the per-operation exchange size. Tuning this parameter sets a tradeoff between per-operation TEE cost and the number of asynchronous synchronizations, in which reducing the latter significantly benefits the overall performance.

The contributions of this paper are summarized as follows:
\begin{packeditemize}
    \item We provide a detailed overview of GPU TEE overheads in inter-GPU communication during DDP training. To ensure secure data transfer, GPU TEE applies AES-GCM encryption and authentication~\cite{nvidiacc}. We quantify the costs of these security operations introduced by GPU TEEs to quantify their impact on performance. 
   
    \item We observe three major factors that cause performance degradation for GPU TEE-protected DDP training. First, a single all-reduce operation is broken down into multiple sub-operations, which all require encryption and decryption when gradients are exchanged. Second, as the number of participating GPUs increases, the number of sub-operations per all-reduce increases linearly, slowing down the training runtime. Third, DDP launches multiple all-reduce operations asynchronously, slicing per-layer gradient into smaller chunks and hiding communication over computation. Although the asynchronous approach has been successful in non-TEE settings, these frequent communications lead to greater encryption and authentication costs when TEE is enabled. We observe that with a scale of four GPU training, total runtime increases by at most 41.64x in GPT-Xlarge and an average of 8.68x compared to DDP training without TEE.  

    \item %\mengyuan{Please rewrite the bullet to something that sounds like: we spend efforts in examing many things, and we then identify something that can help here, instead of directly talking about the solution. }
    We systematically explored various benchmarks and hyperparameter configurations in DDP to mitigate the performance overhead introduced by secure communication in multiple GPU TEEs. Through this investigation, we identify the hyperparameter configuration of the asynchronous all-reduce packet size, controlled by bucket\_cap\_mb, as a key factor. By tuning this parameter to batch multiple asynchronous communications into larger packets, we were able to substantially reduce TEE-induced overheads. Specifically, the optimal bucket size reduces TEE overhead by 4.95x and 7.31x compared to a default setting in two-GPU and four-GPU TEE settings, respectively. Nevertheless, relative to non-TEE DDP training, the two-GPU and four-GPU TEE settings still experienced 1.62x and 3.03x runtime degradation, highlighting future solutions in secure multi-GPU training despite proposed mitigations.
    
\end{packeditemize}

%% file: 3_background.tex
\subsection{Trusted Execution Environment}
%\mengyuan{Talk in general what the main idea behind TEE here. For example, hardware-rooted trust, the problem it wants to solve, and then connect to GPU-TEE}.
With the increasing need for private and secure computation in cloud settings, various hardware-enforced Trusted Execution Environments (TEEs) have been introduced. TEEs are designed to provide an isolated and secure execution environment within the processor with specialized security features implemented in hardware. Trusted Computing Base (TCB) prevents unauthorized access (virtual memory translation requests) or tampering of sensitive data and computation, even from privileged software, such as operating systems or hypervisors. Additionally, DRAM encryption and authentication safeguards against potential hardware attacks such as memory bus probing and data replay attacks. 

Previously, CPU TEEs were more commonly deployed: Intel Software Guard Extensions (SGX)~\cite{sgx}, ARM TrustZone~\cite{trustzon}, AMD Secure Encrypted Virtualization (SEV)~\cite{amdcc}, and Intel Trusted Domain Extensions (TDX)~\cite{tdx}. With the growing need for secure ML workloads, NVIDIA Confidential Computing (CC)~\cite{nvidiacc, nvidiappcie} has extended TEEs to GPUs. This paper focuses on NVIDIA CC.

\subsection{GPU-based TEE}
\label{subsec:gputee}
NVIDIA CC includes two major components: 1) a CPU TEE and 2) a GPU TEE. The CPU TEE system can be an Intel chip with Intel TDX functionality that can provide hardware-enforced program isolation for a virtual machine (VM), namely a confidential VM (CVM). The GPU TEE can be an NVIDIA GPU supporting NVIDIA CC, such as H100. GPU TEEs also provide similar program isolation for processes running on GPUs. In NVIDIA CC, GPU drivers required for GPU TEEs are loaded into the trusted memory of the CPU TEE. Then, CPU TEEs will issue commands to the GPU TEEs through the secure driver.

\bheading{GPU TEE's Threat model.}
\label{sec:threat}
%\mengyuan{we are not attacking something, so maybe place threat model here makes more senses. Unless we have security anlaysis, otherwise we dont need to specify threat model. Help me rewrite the text to fit better here. }
As we do not modify the original security guarantees of the NVIDIA CC framework~\cite{nvidiacc}, we apply the same threat model in our work.
In this threat model, the adversary can take control of the entire host-side software, including the operating system, and probe the physical buses in the cloud system. Thus, the adversary can read/write traffic on the PCIe and NVLink through privileged OS or physical probing. We assume data and code within CPU and GPU packages are secure, including the GPU driver located in the CPU TEE's TCB. We also trust the GPU memory, which is closely placed inside the package and is resistant to external tampering. Similarly to NVIDIA CC, we consider side-channel attacks out of scope.

\input{wrappers/aesgcm}

\bheading{Encrypted Data Transfer.} As the system design trusts only the CPU and GPU hardware packages, all other hardware components are considered untrusted. Thus, interconnections connecting CPU TEEs and GPU TEEs are also untrusted: PCIe (CPU-to-GPU communication link) and NVLink (GPU-to-GPU communication link). To ensure secure transmission, traffic on these links must be encrypted for confidentiality and authenticated for integrity. %This requirement is particularly critical for distributed ML training, where GPUs must synchronize gradients. GPU TEEs must encrypt and generate MAC tags for their gradients before sending them over NVLinks. 
%The next section will discuss the encryption and authentication scheme used in the NVIDIA CC.

In NVIDIA CC, AES-GCM ensures a high level of security for traffic outside the TEE, which happens through PCIe and NVLink. Figure~\ref{fig:aesgcm} demonstrates the AES-GCM algorithm, composed of AES-CTR (AES counter mode), providing confidentiality, and Galois message authentication code (GMAC), generating integrity tags. AES-CTR in Figure~\ref{fig:aesctr} is highly parallel as each AES block cipher is independent of other blocks, making AES-CTR a suitable encryption scheme to be used in GPU parallel computation. However, GMAC tag computation, shown in Figure~\ref{fig:ghash}, is a serialized process, in which the previous Galois field multiplication output is used as input to the next level GF multiplication. Despite its robust security, GMAC's inherent sequential property limits AES-GCM from exploiting GPU parallel processing.

\subsection{Distributed Data Parallelism} 
\label{subsec:back_ddp}
PyTorch Distributed Library~\cite{pytorchddp} is a robust framework designed to efficiently scale machine learning training across multiple devices (e.g. CPUs or GPUs), multiple compute nodes, or a combination of both. Without specific user intervention, the library manages the synchronization of model parameters and gradients across devices while distributing workloads to maximize hardware utilization. PyTorch supports various communication backends, including NVIDIA Collective Communications Library (NCCL)~\cite{nccl}, Gloo, and Message Passing Interface (MPI), allowing the training process to efficiently adapt different system configurations and scales. 

DDP~\cite{pytorchddp} replicates model parameters across all participating devices, with each process running training with a distinct subset of the dataset. Since every device has the same model parameters, their computed gradients must be synchronized to maintain consistency in model parameters at each training iteration. To maximize inter-GPU bandwidth provided by NVLink, all-reduce operations are launched asynchronously instead of synchronizing gradients at the end of each training iteration. Thus, during the backward computation, the communication overhead incurred by all-reduce operations is overlapped with computation. DDP is widely used for its ease of use with minimal code modification, scalability, and broad applicability to various hardware configurations. However, DDP has a key limitation as models must fit inside the memory of each participating device.

\subsection{Ring All-Reduce}
\label{subsec:ringar}

NCCL~\cite{nccl} is NVIDIA's communication framework, optimized to enable data exchange between multiple GPUs with high bandwidth in both a single node and multi-node environments. NCCL primarily focuses on machine learning and high-performance computing (HPC) workloads, supporting various communication operations, such as all-reduce, reduce-scatter, all-gather, and broadcast. NCCL uses ring all-reduce as its default method for the all-reduce algorithm in data-parallel training. 

\begin{figure}
    \centering
    \includegraphics[width=3.5in]{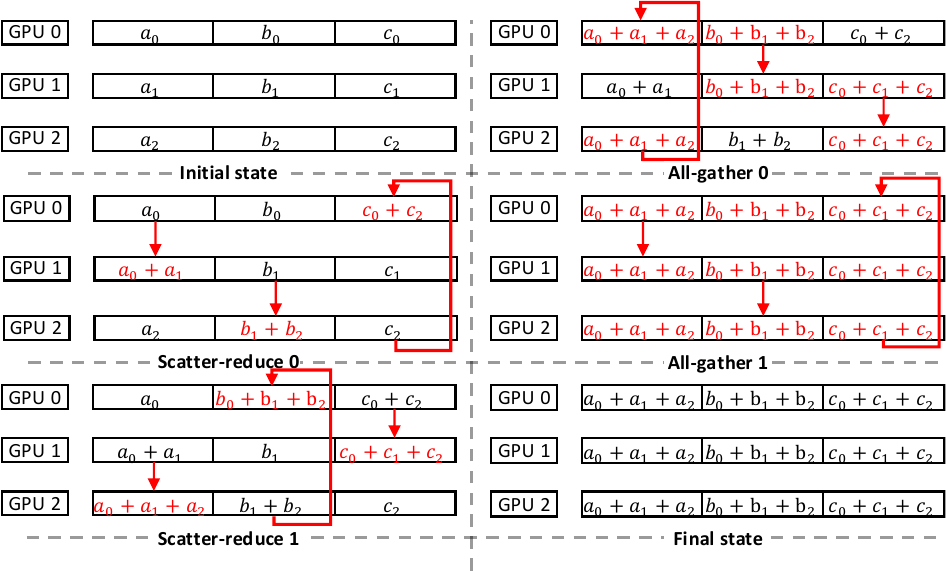}
    \caption{Baseline ring all-reduce algorithm in three GPUs. All-reduce operation in three GPU settings consist of two scatter-reduce and all-gather steps (0 \& 1).}
    \label{fig:ringallreduce}
\end{figure}

Figure~\ref{fig:ringallreduce} demonstrates the ring all-reduce of three GPUs. GPUs are organized in a logical ring topology. Ring all-reduce operates in two phases: scatter-reduce and all-gather. In the \textbf{scatter-reduce} phase, each GPU divides its gradients into chunks $GPU_0(a_0, b_0, c_0)$,$GPU_1(a_1, b_1, c_1)$,$GPU_2(a_2, b_2, c_2)$. Each GPU starts passing its unique chunk around the ring, where each GPU adds its local contribution to the chunk it receives. By the end of this phase, each GPU holds the reduced sum of one chunk of the gradient $GPU_0(b_0+b_1+b_2)$, $GPU_1(c_0+c_1+c_2)$, $GPU_2(a_0+a_1+a_2)$. In the \textbf{all-gather} phase, each GPU distributes the reduced chunks around the ring as demonstrated. After this phase, each GPU has the complete reduced gradient $(a_0+a_1+a_2)$+$(b_0+b_1+b_2)$+$(c_0+c_1+c_2)$, completing the gradient synchronization across all participating GPUs. The number of scatter-reduce and all-gather operations scales linearly with the increase in the number of GPUs. 

%% file: wrappers/aesgcm.tex
\begin{figure}[htbp]
    \centering
    % Left subfigure (a)
    \begin{subfigure}{0.49\columnwidth}
        \centering
        \includegraphics[width=\linewidth]{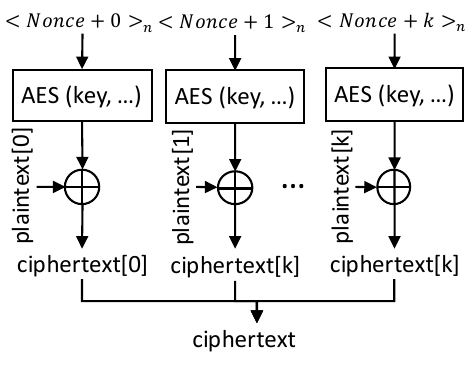}
        \caption{}
        \label{fig:aesctr}
    \end{subfigure}
    \hfill
    % Right subfigure (b)
    \begin{subfigure}{0.49\columnwidth}
        \centering
        \includegraphics[width=\linewidth]{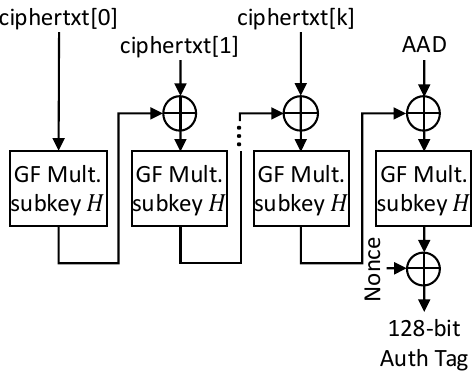}
        \caption{}
        \label{fig:ghash}
    \end{subfigure}
    \caption{Overview of AES-GCM regarding (a) AES-CTR mode to encrypt k blocks of plaintext and (b) GMAC operation to generation MAC tag generation.}
    \label{fig:aesgcm}
\end{figure}

%% file: 4_overview.tex
In this section, we provide an overview of TEE overheads during DDP training. First, we provide an overview of the computation flow in DDP training with AES-GCM encryption and authentication enforced by TEE. Then we go over three primary contributors to performance degradation.  %\mengyuan{apply the same strategy as what I mentioned in contribution here. }
After thorough examination of different DDP hyperparameter settings to address the performance penalties of secure communication in multi-GPU TEEs, we recognized that tuning the asynchronous all-reduce packet size, governed by bucket\_cap\_mb parameter, can greatly reduce the overheads. %In further sections, we use CC interchangeably with TEE-enabled environment.

\begin{figure*}[htbp]
    \centering
    \includegraphics{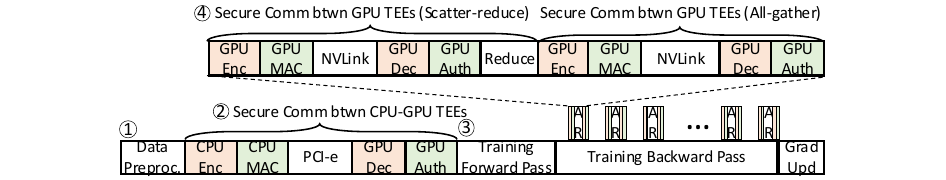}
    \caption{Computation flow of DDP training of each GPU. All-reduce (AR) operation is launched asynchronously, overlapping with training backward pass. Enc/Dec (orange) and MAC/Auth (green) are additional costs to establish secure channel in CPU-GPU and GPU-GPU communication in TEE.}
    \label{fig:computationflow}
\end{figure*}

\subsection{Overview of DDP Training with TEE}
DDP training process in multi-GPU follows the steps outlined in Figure~\ref{fig:computationflow}. The bottom figure shows the overall training flow taken by each GPU, and the top figure shows the flow of a single all-reduce operation, which is one of the sub-operations performed during the back propagation phase. 
The training takes the following steps: (1) the CPU TEE groups input data into batches and performs encryption and MAC tag generation; (2) the CPU TEE sends the encrypted data through PCIe, where the GPU TEEs receive the encrypted data and performs decryption and tag verification for training; (3) each GPU TEE performs the training process; (4) during backward propagation, the GPU TEEs exchange gradients (via all-reduce) over NVLink, encrypting and authenticating each data transfer. As both PCIe and NVLink used in steps 2 and 4 are not trusted, data that leaves the trusted boundary (outside CPU and GPU packages) must be encrypted/decrypted and authenticated.

To enable secure communication in steps 2 and 4, GPU TEEs must first establish a secure channel between CPU TEE and peer GPU TEEs by sharing a symmetric key~\cite{mohan2024securing, zhu2024confidential, wang2024fastrack}. Upon having a shared key between TEEs, the outgoing transmission is encrypted, and a MAC tag is generated using  AES-GCM, which guarantees confidentiality and integrity. Upon receiving the incoming transmission, the encrypted data is decrypted and authenticated. While CPU encryption in step 2 becomes the bottleneck in a single GPU training scenario~\cite{nvidiacc, wang2024fastrack}, such CPU-to-GPU transfer occurs infrequently. Thus, our focus in this paper is GPU-to-GPU gradient exchange in step 4 and its associated TEE overhead, which occurs more frequently in multi-GPU training scenarios, making it the dominant factor of the performance degradation. % Before transmission, the sender GPU TEE must encrypt the data and generate a GMAC tag. The encrypted data and its tag are placed in the bounce buffer for transmission over the link. On the receiving end, the receiver GPU TEE decrypts the encrypted data and verifies the integrity of the received data using the MAC tag.

In the current PyTorch DDP deployment, all-reduce operations (step 4) can occur asynchronously because gradient aggregation is an element-wise operation and thus, can be transmitted eagerly to overlap with backward computation. However, in CC-on setting, each GPU must pay the encryption cost for the outgoing gradients and the decryption cost for incoming gradients. %While all-reduce operations in the figure~\ref{fig:computationflow} are overlapped with backward computation only for demonstration purpose, our evaluations shows that TEE costs that occur between GPU synchronizations are non-negligible. 
We will explain the main factors that contribute to inter-GPU TEE overheads in the following sections.

\subsection{TEE Overhead over Single Ring All-Reduce}
\label{subsec:overview_1}

\input{wrappers/tee_allreduce}

In DDP ML training, all participating GPUs synchronize their model gradients using all-reduce operations at every training iteration, which by default uses ring-based logical topology (ring all-reduce). %\mengyuan{is this correct?} 
These ring all-reduce operations require secure communication between GPU TEEs, introducing additional overhead due to the AES-GCM encryption and authentication. As described in the section ~\ref{subsec:back_ddp}, each ring all-reduce operation is divided into multiple scatter-reduce and all-gather operations. Similar to Figure~\ref{fig:ringallreduce} that shows the all-reduce without TEEs, Figure~\ref{fig:tee_allreduce} shows the general flow of the all-reduce operation among three GPU TEEs. To ensure secure inter-GPU communication, each send/receive within all-reduce and scatter-reduce will be accompanied by encryption/decryption and authentication overhead, as highlighted in the figure.  

%\mengyuan{This is not a standard way. consider changing to something like:} 

\bheading{Scatter-Reduce Overhead.} Scatter-reduce operation involves dividing the gradient into chunks, which are passed between GPUs in a ring topology for summation. The operation involves encrypting and generating a MAC tag from the sender, which is then decrypted and authenticated at each operation step, as every participating GPU must sum up its portion of the gradient to be updated. Figure~\ref{fig:tee_sctred} demonstrates AES-GCM performed at step 0 of the scatter-reduce operation. In scatter-reduce step 0, GPU0 encrypts and generates a tag for its chunk, $a_0$, as the sender, while GPU1 decrypts and authenticates $a_0$ as the receiver and sums it with its own chunk $a_1$. A similar procedure happens for GPU1 and GPU regarding $b_1$, and GPU2 and GPU0 regarding $c_2$. Compared to scatter-reduce explained in Figure~\ref{fig:ringallreduce}, additional encryption and tag generation overhead as a sender and decryption and authentication overhead as a receiver are added into the pipeline. As encryption and decryption, tag generation and authentication are essentially the same operations, secure communication between two GPU TEEs results in two encryption and authentication costs per GPU.

\bheading{All-Gather Overhead.}
%\replaced{GPU TEE also}{we must} 
Similarly, GPU TEE also adds two encryption and authentication costs to all-gather operation as shown in Figure~\ref{fig:tee_allgat_it0}. During all-gather, GPU2 sends its reduced gradient chunk ($a_0+a_1+a_2$) to GPU0 and GPU1. GPUs in the middle of the ring (GPU0 for gradient chunk $a_0+a_1+a_2$) must decrypt the received chunk, re-encrypt it, and forward it to the following GPU. Figure~\ref{fig:tee_allreduce} highlights how scatter-reduce and all-gather phases in ring all-reduce contribute to significant GPU TEE overhead due to secure communication requirements. 

\subsection{Additional Scaling Factors of All-Reduce}
\label{subsec:overview_2}
We identify two additional factors that exacerbate the additional cost of GPU TEEs in DDP: the number of participating GPUs and the model size. 

%\para{Number of participating GPUs} 
%\subsection{Number of Participating GPUs}
%\mengyuan{The subsection title doesnt make senses. reader should know what the subsection is about by reading the title.}
\bheading{Number of Participating GPUs.}
TEE overhead in ring all-reduce operation grows linear to the number of GPU TEEs involved in DDP training. With increased number of GPUs, gradients calculated in backpropagation per GPU are partitioned into smaller chunks, which require more sub-operation steps for a single all-reduce operation. For $n$ GPU TEEs, each ring all-reduce operation consists of $n-1$ scatter-reduce and $n-1$ all-gather steps. During each step, a GPU performs one encryption (including MAC generation) as well as one decryption (including MAC authentication), resulting in a total of $4\times(n-1)$ encryption and authentication overhead for every ring-all-reduce operation.

\begin{figure}
    \centering
    \includegraphics[width=\columnwidth]{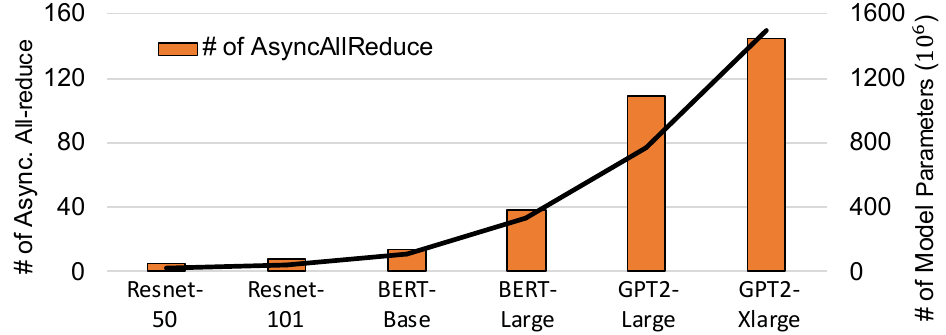}
    \caption{Number of model parameters (line) and number of asynchronous all-reduce operations per training iteration (bar).}
    \label{fig:paramasync}
\end{figure}

%\para{Model size} 
%\subsection{Trained ML Model Size}
%\mengyuan{The same}
\bheading{Training Model Size.}
Apart from the number of GPUs participating in DDP, the second factor is the training model size. Figure~\ref{fig:paramasync} represents the relationship between model size and the number of asynchronous all-reduces at each training iteration, based on profiling results conducted through NVIDIA Nsight Systems. As model size increases, asynchronous all-reduce operations increase proportionally, ranging from five all-reduces in Resnet50 to 142 all-reduces in GPT2-Xlarge. The reason for increased all-reduce operations with model size is that larger model sizes split the data into many more fixed-size chunks and transmit them while overlapping the data transmission with the computation of another chunk of gradients.  

%Without TEE, these asynchronous operations take at most 2.6\% of the backward propagation execution time in GPT2-Xlarge, effectively hiding the communication overhead within the computation. However, the use of GPU TEEs significantly amplifies this communication due to the required encryption and authentication. 
Without TEE, these asynchronous operations during the backward propagation take at most 2.6\% of the entire  GPT2-Xlarge, effectively hiding the communication overhead within the computation. However, the use of GPU TEEs significantly amplifies this communication due to the required encryption and authentication. 
With $k$ asynchronous all-reduces, the total number of encryption and authentication is $4\times k \times (n-1)$, as shown in subsection~\ref{sec:overview}. For instance, this overhead ranges from $4\times5\times(4-1)=60$ in Resnet50 up to $4\times142\times(4-1)=1704$ in GPT2-Xlarge in $n=4$ GPU TEEs. %In Section~\ref{sec:experiment}, our experiments reveal how these overheads dominate training runtime, making it impractical to fully hide them under training backward computation time. 

\subsection{Tuning DDP Gradient Synchronization Size}
%\mengyuan{This is a huge weaken way for the paper. It is supposed to be one of our major contributions. But now it looks identical as section 3.4, reader may not even realize this is a potential optimization. We need to highlight our effort before drawing the conclusion (e.g., changing capmb can improve performance) } 
%\mengyuan{Consider mergae 3.3-3.4 into one subsection , and make 3.5 the insight/conclusion draw from 3.3-3.4 analysis. Or other way to highlight 3.5}
The analysis in Section~\ref{subsec:overview_2} reveals that TEE overhead in DDP training grows significantly with both the number of participating GPUs and the number of asynchronous all-reduce operations, which in turn scales with the model size. In particular, the encryption and authentication costs accumulate rapidly, creating a major performance bottleneck. 

To address this, we explore an optimization opportunity in PyTorch DDP that controls the frequency of asynchronous all-reduce launches via bucket\_cap\_mb hyperparameter.
%PyTorch DDP launches asynchronous inter-GPU communication during training, based on hyperparameter, bucket\_cap\_mb. 
Depending on the model size, the bucket can be filled once or multiple times within a single layer, in which each filled bucket launches an asynchronous all-reduce. By tuning the bucket size, we can batch multiple gradient chunks into a single all-reduce, reducing the number of secure communication steps and thereby mitigating TEE-induced overhead.% We use this hyperparameter to batch the asynchronous all-reduce operations and reduce the TEE overheads. We will demonstrate the effectiveness of batching asynchronous all-reduce in the following Section.

In Section~\ref{sec:experiment}, we evaluate the TEE overheads in the overall training runtime, and quantify how secure communication  can impact larger models that require frequent gradient exchange. Additionally, we evaluate the impact of our proposed mitigation strategy to evaluate whether multi-GPU DDP training in TEEs still experiences  performance degradation.

%% file: wrappers/tee_allreduce.tex
\begin{figure}[htbp]
    \centering
    % Left subfigure (a)
    %\begin{subfigure}{0.49\columnwidth}
    \begin{subfigure}{0.49\columnwidth}
        \centering
        \includegraphics[width=\linewidth]{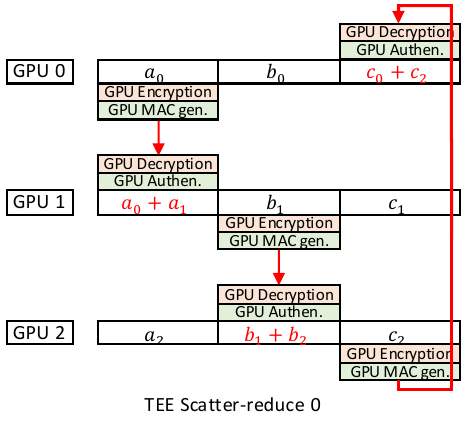}
        \caption{}
        \label{fig:tee_sctred}
    \end{subfigure}
    \hfill
     % Right subfigure (b)
    \begin{subfigure}{0.49\columnwidth}
        \centering
        \includegraphics[width=\linewidth]{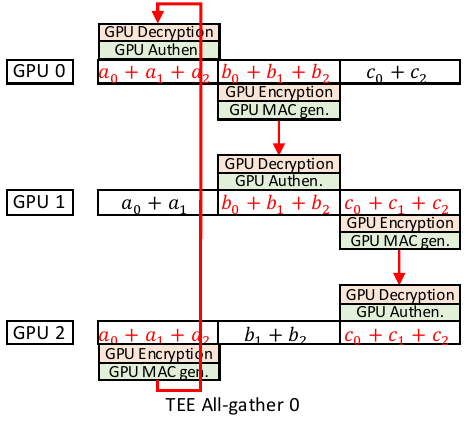}
        \caption{}
        \label{fig:tee_allgat_it0}
    \end{subfigure}
    \hfill
    \caption{Ring all-reduce operation using AES-GCM in multiple GPU TEEs. (a) demonstrates the first operation step of scatter-reduce operation in multiple GPU TEEs. (b) demonstrates the first operation step of all-gather operation in multiple GPU TEEs. Number of operation steps differ by the number of participating GPUs.}
    \label{fig:tee_allreduce}
\end{figure}

%% file: 5_methodology.tex
\subsection{Experimental Setup}

%\mengyuan{section 4 should have a meaningful title. .Actually 4.1 sounds more like experiment setup. }

\begin{table}[ht]
  \centering
  \renewcommand{\arraystretch}{1.2} 
  \caption{Overview of the evaluated models.}
  \label{table:bench}
  \begin{tabular*}{\columnwidth}{@{\extracolsep{\fill}} l l l r}
    \toprule
        % \midrule
    Benchmark   & Category         & Dataset                    & Batch Size \\ 
    \midrule
    ResNet-50   & CNN              & ImageNet~\cite{imagenet}   & 1024       \\
    ResNet-101  & CNN              & ImageNet~\cite{imagenet}   & 1024       \\
    BERT-Base   & NLP (Encoder)    & GLUE~\cite{glue}           & 128        \\
    BERT-Large  & NLP (Encoder)    & GLUE~\cite{glue}           & 128        \\
    GPT2-Large  & LLM (Decoder)    & CrowS-Pairs~\cite{crows}   & 1          \\
    GPT2-XL     & LLM (Decoder)    & CrowS-Pairs~\cite{crows}   & 1          \\
    \bottomrule
  \end{tabular*}
\end{table}

We implemented representative ML models, shown in Table~\ref{table:bench}, to characterize the overheads of GPU CC-based DDP training. We use Resnet~\cite{resnet} to represent the convolution network, particularly Resnet50 and Resnet101 from PyTorch repository~\cite{pytorchex}. BERT~\cite{bert} and GPT2~\cite{gpt2} represent transformer-based encoder and decoder language models used in natural language processing, in which we use Huggingface Accelerate repository~\cite{huggingface,accelerate}. %GraphSAGE from PyTorch Geometric repository~\cite{pytgeo} is used to represent GCN models.

Before presenting our experimental results, we first explain the experimental setup used to evaluate the multiple-GPU TEE system. While NVIDIA describes all the protocols they use for distributed training in their documentation~\cite{nvidiacc}, they have not publicly released the compatible drivers that support distributed CC training under a single CVM. In order to provide quantitative data and attribute the overheads to various steps within the DDP training, we implemented the security primitives for encryption, decryption, and MAC authentication as GPU CUDA kernels that perform the same security tasks as would have been done within the NVIDIA driver. For CPU-side encryption and tag generation, we use OpenSSL library~\cite{openssl}.

\begin{figure}
    \centering
    \includegraphics[width=\columnwidth]{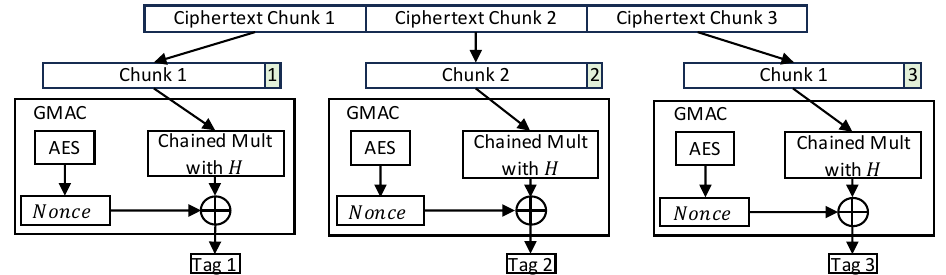}
    \caption{Multi-chaining authentication with ciphertext divided into three chunks.}
    \label{fig:multichain}
\end{figure}

%\mengyuan{Think as the reader, the reader will have the question: Why the author mention that they we will use this authentication scheme, why not directly use Nvidia's way. Without a motivation, it then become weird. Actually, reviewer will for sure ask the result for the unmodified Nvidia's performance. If this the case, how can we rewrite the paragraph to address reader's concerns, or do things in a more natural way. } \mengyuan{If the first part of section 4.2 is the answer to this, we should put them togther: Directly provide the answer before the reader thinking about the qeuestion. }
For GPU encryption/decryption and MAC generation and authentication, NVIDIA hardware may employ proprietary cryptographic accelerators in their machine. However, as their detailed implementation is not publicly disclosed, we use custom-built CUDA kernels and validate our setup in Section~\ref{subsec:validation}. Note that no current GPU TEE supports a performance-boosting feature called Trusted I/O (TIO); therefore, we do not include TIO boosting in our measurements with more details provided in Section.\ref{sec:related}.
Compared to AES-CTR encryption, which is highly parallel and can exploit GPU parallelism even with custom kernel, GHASH-based authentication is innately serial process as described in Figure~\ref{fig:aesgcm}.% Section~\ref{subsec:aesgcm}.
Furthermore, given that inter-GPU data transfer during gradient synchronization in models such as GPT2 can reach up to 6GB, implementing a strictly serial authentication kernel would severely degrade performance and fail to accurately reflect realistic system behavior. Therefore, we adopt the multi-chaining authentication scheme, presented in ~\cite{wang2024fastrack} to improve parallelism in GPU-based MAC generation and authentication. 
%As described in Figure~\ref{fig:ghash}, AES-GCM MAC generation and authentication is inherently a serial process. Previously, a single long chain of block multiplication is used to generate a MAC tag for an input. Fastrack~\cite{wang2024fastrack} proposes multi-chaining authentication illustrated in Figure~\ref{fig:multichain}, which divides this single chain into multiple chunks. 
Illustrated in Figure~\ref{fig:multichain}, multi-chaining authentication, divides a long chain of GHASH into multiple small chunks. Each chunk, attached with additional associated data (AAD) to indicate the chunk order, is processed through different GMAC blocks, enhancing parallelism. %We apply multi-chaining authentication in GPU-side authentication including both CPU-to-GPU transfer and inter-GPU transfer. 
%As the CPU-to-GPU transfer size is 190x larger in GPU-to-GPU communication, we use two-byte AAD to represent the order of 65536 chains, which adds negligible GPU authentication overhead to the total runtime.
We use two-byte AAD to represent the order of 65536 chains, which adds negligible GPU authentication overhead to the total runtime.
 
\subsection{Setup Validation}
\label{subsec:validation}

Before embarking on the multi-GPU characterization, we setup an single GPU TEE using NVIDIA H100 GPU and Intel Xeon Gold 6548Y$+$ paired with 512 GB of memory for our CPU setup to validate our emulation. We used OS kernel version of 6.2.0-mvp10v1+8-generic and NVIDIA Driver version of 550.54.13 with CUDA Toolkit 12.4. The VMs were configured as 32-virtual CPUs (vCPUs), 64-GB memory, and 100-GB storage for CC-on and CC-off. We compare our baseline of a single GPU TEE implementation with a physical CC on NVIDIA H100. 
%\mengyuan{list about software version here as well}
%\mengyuan{check section 3.2 in https://arxiv.org/pdf/2403.03360}

\begin{figure}
    \centering
    \includegraphics[width=\columnwidth]{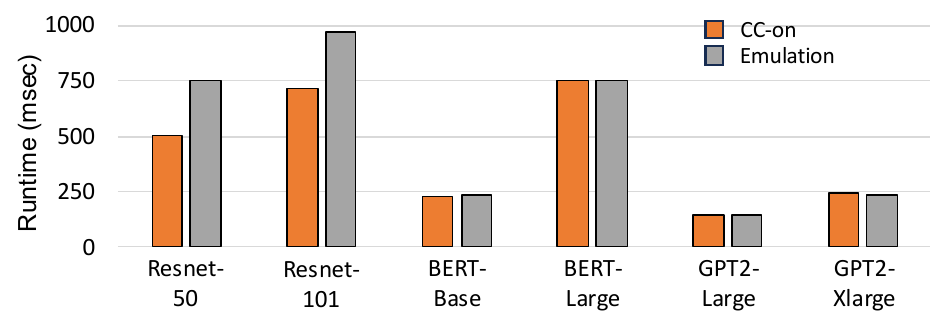}
    \caption{Validation of AES-GCM implementation on a single GPU training.}
    \label{fig:validation}
\end{figure}

Figure~\ref{fig:validation} shows the single epoch training runtime measured in the GPU TEE and our implementation of CPU/GPU AES-GCM. Resnet50 and Resnet100 implementation using our kernels are within 16\% and 13\%, of the native GPU TEE implementation results that we obtained. Transformer-based BERT and GPT2 models show an average of 1.3\% divergence in our implementation compared to the actual GPU TEE. The performance discrepancy between the actual measurement and our implementation comes from the difference in CPU AES-GCM throughput. While NVIDIA claims to achieve 4GB/s bandwidth for CPU AES-GCM~\cite{nvidiacc} without further disclosure of their implementation strategies, our implementation using OpenSSL only achieves about 1.82GB/s. However, such performance difference is not critical in observing TEE overhead in multi-GPU training, as CPU-to-GPU data transfer only happens once at the beginning of training.

In the following experiments, we analyze the inter-GPU communication overhead of GPU TEEs per training iteration with a fixed batch size. The size of gradient updates during all-reduce in DDP back-propagation is determined by the model size rather than the batch size. A per-batch analysis would demonstrate that the TEE overhead of GPU-to-GPU transfer creates a system bottleneck. Thus, our experiments analyze workloads per training iteration with fixed batch sizes to provide an accurate assessment of inter-GPU communication overheads. 

%% file: 6_evaluation.tex
In this section, we characterize the overheads introduced by GPU TEEs in DDP training. Section~\ref{subsec:2gpu} and ~\ref{subsec:4gpu} present a detailed runtime analysis of two-GPU and four-GPU training, highlighting the associated TEE overheads. We observe that the three factors, as discussed in Section~\ref{sec:overview}, significantly impact the TEE overheads and degrade performance. In Section~\ref{subsec:batch}, we validate the mitigation technique of tuning bucket\_cap\_mb. Lastly
After thorough examination of different DDP hyperparameter settings to address the performance penalties of secure communication in multi-GPU TEEs, we recognized that tuning the asynchronous all-reduce packet, governed by bucket\_cap\_mb, can greatly reduce the overheads. Lastly, we discuss the impact of TEE overheads when we scale DDP training to 8 GPUs and more.
%: breakdown of all-reduce operations into multiple sub-operations, number of GPUs participating, and number of asynchronous gradient exchange determined by the model size. 
%Our observations, as discussed in section~\ref{sec:overview}, indicate that both factors, the number of GPUs and the number of asynchronous all-reduce, contribute substantially to encryption and authentication overhead in DDP training. Section~\ref{subsec:batch} demonstrates the performance benefit of increasing the communication size per asynchronous all-reduce operation by manually tuning DDP hyperparameter, bucket\_cap\_mb. In section~\ref{subsec:discuss}, we discuss the superlinear performance degradation of increasing the number of GPUs beyong 8 GPUs and the CC overhead of different parallelism schemes used in ML community. 

\input{wrappers/2gpuExecDecomp}
\subsection{Two GPU TEE Performance Analysis}
\label{subsec:2gpu}
Figure~\ref{fig:2gpuexec} compares the per-training-iteration runtime of DDP training with CC-off and CC-on using two GPUs. %With CC-off, the per-iteration runtime ranges from 20 msec for GraphSAGE to 800 msec for BERT-Large on each GPU TEE. 
%\mengyuan{maybe provide a short insight here, and then discuss details.}
We observe that enabling CC dramatically increases per-iteration runtime with TEE overheads in inter-GPU communication dominating workloads, even in applications with less frequent all-reduce operations like ResNet. However, such performance degradation is more dominant in Bert and GPT2 models with frequent all-reduce operations.
With CC-off, the per-iteration runtime ranges from 200 msec for Resnet50 to 800 msec for BERT-Large on each GPU TEE. Although BERT-Large has one-fifth the model size of GPT2-Xlarge, the number of per\_device\_train\_batch\_size is greatly limited in GPT2-Xlarge to only a single batch due to GPU memory capacity limitation. 

%When CC is enabled, the per-iteration runtime increases significantly, ranging from 1.61x in GraphSAGE to 16.78x in GPT2-Large. 
When CC is enabled, the per-iteration runtime increases significantly, ranging from 1.97x in Resnet50 to 16.78x in GPT2-Large. This increase is caused by AES-GCM encryption and authentication overhead of asynchronous all-reduce operations. For instance, compared to BERT-Large, which has 38 asynchronous all-reduces, GPT2-Xlarge executes 142, nearly four times more, resulting in a proportional increase in inter-GPU TEE overhead. 

%The overhead of TEE can differ depending on the batch size. As with a larger batch size per device, CC-off runtime can take up a large portion of CC-on runtime, as the number of asynchronous all-reduce operations and their size are fixed to the model size. However, 

Figure~\ref{fig:2gpudecomp} demonstrates the runtime decomposition of CC-on with two GPUs. TEE overheads dominate the runtime of applications with a larger number of asynchronous all-reduce. For example,  TEE overheads occupy over 93\% of total runtime in both GPT2 workloads. In contrast, training computation in GPU, which consists of forward and backward propagation, contributes to an average of 40\% and 52\%  of runtime in BERT and Resnet models, respectively. BERT and Resnet models have smaller models compared to GPT2, experiencing fewer asynchronous all-reduces. Also, as these models allow larger batch training, their training computation portion is larger than that of GPT2. 

Notably, in Resnet models, CPU encryption and authentication overhead accounts for 6.2\% in Resnet101 and 8.6\% in Resnet50. This is because training image with a batch size of 1024 must be encrypted and authenticated to be transferred to each GPU TEEs. 

%GraphSAGE suffers the least TEE overhead with only 33.5\%, as it executes only two all-reduces over its backward computation. A significant portion of PCIe transfer time in GraphSAGE, accounting for 14.7\%, i  attributed to data preprocessing time, which we have included in the PCIe transfer measurement for clarity in our figures. While data preprocessing is relatively minor for CNNs and language models, graph data preprocessing using NeighborLoaders incurs substantial time per training iteration.

\subsection{Four GPU TEE Performance Analysis}
\label{subsec:4gpu}
\input{wrappers/4gpuExecDecomp}
%\mengyuan{maybe provide a short insight here in the beginning somewhere, and then discuss details.}
Compared to the two-GPU setting, the four-GPU settings shows larger runtime slowdown with increased TEE overheads due to increase in the number of participating GPUs as discussed in Section~\ref{subsec:overview_2}.
Figure~\ref{fig:4gpuexec} compares the training runtime four GPUs. For CC-off, the runtime of four GPUs is, on average, 1.2x longer than two-GPU due to an increase in PCIe time. Forward and backward computation time remains consistent per device since the per-device batch size is fixed. However, the total training batch size over the machine doubles, resulting in an increase in PCIe time. NVLink communication cost of gradient synchronization, which includes reduce operations at each GPU and transfer time over the NVLink, remains unchanged relative to PCIe because the cost of NVLink is determined by the size of the model. %With CC-on, the training iteration runtime increases by 1.85x in GraphSAGE to 42.36x in GPT2-Large. 
With CC-on, the training iteration runtime increases by 3.29x in Resnet50 to 42.36x in GPT2-Large. Similar to the two GPU cases, the TEE encryption and authentication regarding inter-GPU communication overhead are the main contributors to this performance degradation. While the number of asynchronous all-reduce remains the same, the number of encryption and authentication triples due to increase in the number of participating device.

The runtime decomposition, shown in Figure~\ref{fig:4gpudecomp}, indicates that the percentage of TEE overhead in total runtime increases from a minimum of 4.5\% in GPT2-Xlarge to a striking 75.1\% in Resnet101 compared to two GPU cases in Figure~\ref{fig:2gpudecomp}. With four GPU TEEs, the TEE overhead triples compared to two GPU TEEs due to a proportional 3x increase in both scatter-reduce and all-gather operations described in Section~\ref{subsec:back_ddp}. Particularly, the runtime of both Resnet models, Resnet50 and Resnet101, increases by 72.4\% and 75.1\%. Compared to two GPU CC-on setup where TEE overhead accounts for less than 40\% of total runtime, the TEE overhead becomes a critical factor, dominating the total runtime in four GPU setup.

%TEE overhead regarding batched input transfer over PCIe accounts for only 1.4\% and 5.6\% in total runtime in two GPU and four GPU experiments, respectively. However, models that train with large batch sizes, such as GraphSAGE and Resnet with 1024 batches, show that CPU-GPU communication overhead accounts for 8.0\%, 11.1\%, and 19.0\% in Resnet50, Resnet101, and GraphSAGE. The transferred byte sizes for these models in CPU-GPU communication compared to inter-GPU communication are 1.5x and 0.53x. 0.58x in Resnet50, Resnet101, and GraphSAGE, respectively. Smaller transferred byte sizes and less frequent asynchronous all-reduce operations account for a larger portion of TEE's CPU-GPU communication overheads. 

\subsection{Batching Asynchronous All-Reduce}
\label{subsec:batch}
Previous experiments demonstrate critical performance degradations when using the PyTorch DDP Library default configuration for bucket\_cap\_mb, which determines the size and number of asynchronous reductions. While such a default configuration enables better overlap of compute and communication in CC-off settings, which maximizes overall training performance, our work has demonstrated that with CC-on, the number of asynchronous reductions should be limited, even if that means reduced opportunity to overlap compute and communication.

\begin{table}[!ht]
    \centering
    \caption{Training Configuration and Per-Transfer Cost of Different bucket\_cap\_mb.}
    \label{table:batched}
    \resizebox{\columnwidth}{!}{
    \begin{tabular}{|c|c|c|c|}
    \hline
        bucket\_cap\_mb & Reduction Size (MB) & Number of Transfer & Per-Transfer Cost \\ \hline
        Default & 44.0  & 145 & 1\\ \hline
        100 & 123.0 & 49 & 1.08 \\ \hline
        200 & 245.9 & 26 & 1.24\\ \hline
        400 & 430.4 &  9 & 1.60\\ \hline
        800 & 860.7 &  8 & 2.69 \\ \hline
       
    \end{tabular}
    }
\end{table}

\begin{figure} [h]
    \centering
    \includegraphics[width=\columnwidth]{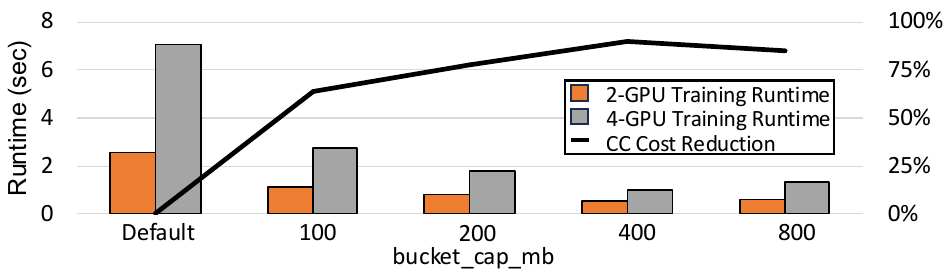}
    \caption{Total runtime and total CC cost reduction with increase of bucket\_cap\_mb.}
    \label{fig:batched}
\end{figure}

The size of asynchronous reduction can be altered using “bucket\_cap\_mb” configuration. By increasing the configuration, we can increase each reduction size, while reducing the number of transfers. We tune “bucket\_cap\_mb” when training GPT2-Xlarge in 2-GPUs to find the best performance.

Table~\ref{table:batched} shows the actual training configuration in different bucket\_cap\_mb. The per-transfer CC cost, which includes the sender's encryption and MAC generation and the receiver's decryption and authentication cost, increases sub-linearly with doubling the bucket\_cap\_mb. As AES-CTR encryption is highly parallel, doubling the encryption size has minimal performance overhead. Authentication cost increases sub-linearly until the multi-chaining authentication scheme exceeds GPU parallelism, at bucket\_cap\_mb is 800, in which the authentication cost doubles. However, compared to the sub-linear increase in per-transfer cost, the number of transfers decreases linearly until bucket\_cap\_mb is 800. % isthere a areason for such stop?

Figure~\ref{fig:batched} shows the total runtime based on Table~\ref{table:batched} in two-GPU and four-GPU settings. With the bucket size configured to 400MB, the total CC cost reduces by 90\%, in which the number of transfers is the main contributor to the reduction in CC cost as described in Section~\ref{sec:overview}. However, the bucket size configuration of 800MB only reduces the all-reduce number by one, while increasing per-transfer cost by 67\%, resulting in only a 50\% increase in total CC cost compared to 400MB.

Bucket size of 400MB shows the best performance in both two- and four-GPU settings, reducing the total runtime by 4.94x and 7.31x compared to the default setting of 44MB with CC-on, respectively. Comparing the runtime to CC-off settings, CC-on runtime increases by only 62\% in the two-GPU setting, which indicates that batching asynchronous all-reduces can bring a significant performance benefit in the CC-on multi-GPU training scenario. The four-GPU scenario shows a 3.04x increase in runtime compared to CC-off with the default bucket size. While the CC overhead exacerbates with an increase in the number of participating GPUs, the 3.04x increase in runtime compared to CC-off with default bucket size is significantly less than the 41.64x increase of CC-on with default bucket size.

\begin{figure} [ht]
    \centering
    \includegraphics[width=\columnwidth]{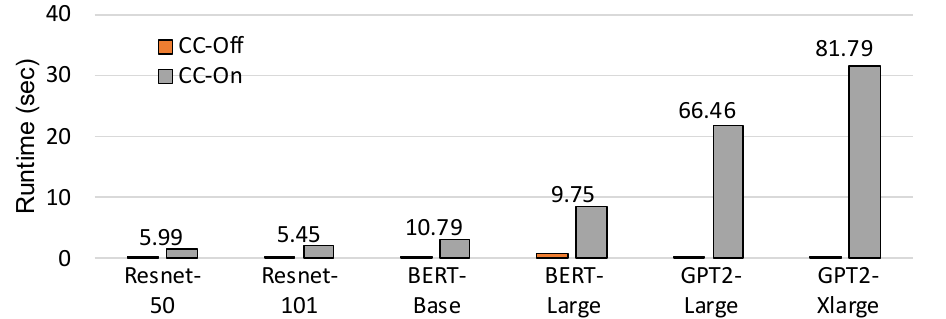}
    \caption{DDP training on 8 GPU TEEs comparing the per-training-iteration runtime (sec) with CC-off and CC-on.}
    \label{fig:8gpuexec}
\end{figure}

\subsubsection{Scaling beyond 8 GPU TEEs}
\label{subsec:scale}

Figure~\ref{fig:8gpuexec} illustrates the runtime of a single training iteration with the per-device batch size in Table~\ref{table:bench}. As 8 GPUs incur a total of 28 encryption and authentication overhead per all-reduce operation, the runtime increase by an average of 16x and striking 81.8x in GPT2-Xlarge, with TEE overhead accounting for 98.8\% of the total runtime per iteration. 

%When scaling beyond eight GPUs, the GPUs are typically distributed across multiple nodes and thus, multiple CVMs must exchange gradients via interconnects other than NVLink. When data passes through a CPU TEE for inter-node gradient synchronization, it must be decrypted, authenticated, and re-encrypted with CPU engine, introducing a major performance degradation as mentioned in the NVIDIA document~\cite{nvidiacc}. Prior works~\cite{wang2024fastrack} also demonstrates major performance overhead in CPU encryption. In multi-node training, CPU's encryption and authentication becomes the system bottleneck. Additionally, as PCIe is approximately 7x slower than NVLink, performance degradation of inter-GPU commmunication across nodes exacerbates, further slowing down the synchronization phase.
When scaling beyond eight GPUs, the GPUs are typically distributed across multiple nodes, where we exchange gradients between multiple CVMs, which exist per node. As data must pass through a CPU TEE for inter-node gradient synchronization via interconnects other than NVLink, the system is introduced with a large number of CPU-CPU and CPU-GPU communications, in which CPU TEE's encryption authentication plays the major system bottleneck as described in NVIDIA document~\cite{nvidiacc}.

\section{Discussion on Distributed Training}
\label{sec:discuss}

%\mengyuan{is this a subsection?}``
\subsection{TEE Overhead in Tree-based All-reduce}
%\mengyuan{Again, what does this title mean. The reader dont know what you want to talk in this subsection.}
NCCL also provides tree-based logical all-reduce, which is implemented with reduction and broadcast operations. Compared to ring all-reduce, which evenly distributes encryption/authentication cost to all the participants, tree all-reduce often burdens the middle-node GPU in its incoming and outgoing bandwidths. As encryption/authentication operate using a shared one-time-pad between each pair of the GPUs in a sequential manner, the middle-node GPUs must decrypt two children's incoming gradients and encrypt one outgoing gradient to its parent node during reduction and vice versa during broadcast. 
% Added implementation Result after IISWC2025
Our implementation results show an average of 1.81x increase in overall runtime compared to ring-based all-reduce because both leaf/root nodes must wait for the middle node to finish its all-reduce stage. Prior works have demonstrated that tree-based reduce operations suffer slowdowns when handling large messages in a small number of nodes due to bandwidth~\cite{cho2023logical}. Our results show that bandwidth is greatly limited in the middle node in a tree-based topology
Thus, even when GPUs use tree all-reduce to exchange gradients, our characterization still holds valid arguments as the components that affect the CC cost (GPU number, model, and exchanged gradient size) are orthogonal to the type of all-reduce operation.

%NCCL also provides tree-based logical all-reduce, implemented with reduction and broadcast operations. However, reduction/broadcast operations burden the middle-node GPUs in their incoming and outgoing bandwidths, which further aggravates the CC cost. As encryption/authentication operate using a shared one-time-pad between each pair of the GPUs in a sequential manner, the middle-node GPUs must decrypt two children's incoming gradients and encrypt one outgoing gradient to its parent node during reduction and vice versa during broadcast. Compared to ring all-reduce, which evenly distributes encryption/authentication cost to all the participants, middle GPUs in tree all-reduce pay 1.5x encryption/authentication cost, slowing down the pipelined reduction/broadcast operation. If we optimize tree all-reduce by overlapping reduction and broadcast operations as proposed in \cite{cho2023logical}, the overhead increases to 3x. Hence, in the CC-on setting, tree-based all-reduce shows impracticality in actual usage. 

\subsection{Different Parallelism Scheme}
\label{subsec:parallel}
With the emergence of larger models, different parallelism techniques, such as pipeline parallelism~\cite{huang2019gpipe} or fully-sharded data parallelism (FSDP)\cite{zhao2023pytorch} have been introduced in machine learning.

%\mengyuan{Change every para to bheading, which makes the paper looks tighter}
%\para{Pipeline Parallelism}
\bheading{Pipeline parallelism}~\cite{huang2019gpipe} splits neural network models into segments of multiple layers and places each segment on a different GPU. Intermediate values are processed in each GPU and passed down to the next segment sequentially via send/receive operations in both forward and backward propagation. Pipeline parallelism is often combined with data parallelism in ML training engines like DeepSpeed~\cite{rasley2020deepspeed}, in which each data parallel rank performs pipeline parallelism, and gradient synchronizations are performed inter-ranks between GPUs that hold the same segments of the model. In such scenarios, our characterization still holds a valid argument as gradient synchronization between different ranks requires a standard all-reduce, which is heavily impacted by the CC overhead.

%\para{Fully-Sharded Data Parallelism} 
\bheading{Fully-sharded data parallelism} (FSDP)~\cite{zhao2023pytorch} minimizes memory overhead in participating devices by sharding model parameters across FSDP units, which consist of multiple devices. Before forward and backward passes, all devices within the FSDP unit must synchronize their parameters through an all-gather operation. Additionally, during the backward pass, the reduce-scatter operation distributes only the necessary gradients and optimizer states to each device, reducing the memory footprint even further. 
% ADDed Results after IISWC2025
For instance, uring FSDP training of GPT-Xlarge, forward pass launches two all-gather operations and backward pass launches one all-gather and two reduce-scatter operations per iteration. In two-GPU implementation, the CC-on total runtime increases by 21\% compared to the CC-off setting. 
As each inter-GPU communication only exchanges shared weights and corresponding gradients, CC cost is smaller compared to the asynchronous all-reduces in DDP, which involve the entire model parameters.

\begin{comment}
\subsubsection{Tensor Parallelism}
Tensor parallelism shards a single layer of a neural network across multiple devices, each device having vertically sharded model. Tensor parallelism enables training of extremely large models, such as GPT-based architectures in which a single layer's memory footprint can exceed the capacity of a single GPU. While taking advantage of the combined memory and compute power of mulitple GPUs, tensor parallelism induce additional communication overheads 
\end{comment}

%% file: wrappers/2gpuExecDecomp.tex
\begin{figure}[t]
    \centering
    % Left subfigure (a)
    \begin{subfigure}{\columnwidth}
        \centering
        \includegraphics[width=\linewidth]{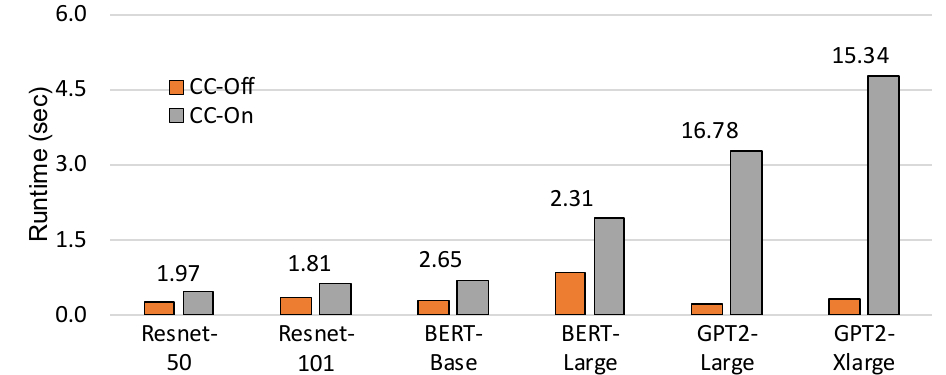}
        \caption{}
        \label{fig:2gpuexec}
    \end{subfigure}
    \hfill
    % Right subfigure (b)
    \begin{subfigure}{\columnwidth}
        \centering
        \includegraphics[width=\linewidth]{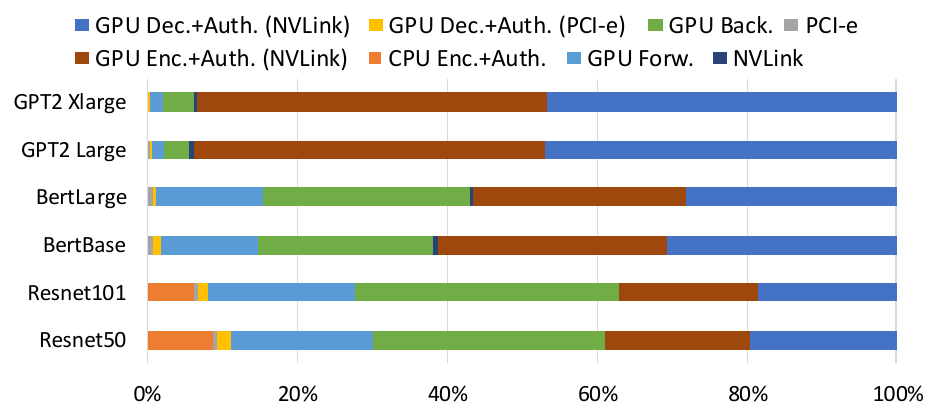}
        \caption{}
        \label{fig:2gpudecomp}
    \end{subfigure}
    \caption{DDP training on 2 GPU TEEs. (a) compares the per-training-iteration runtime (sec) with CC-off and CC-on. (b) demonstrates the runtime decomposition of 2 GPU training with CC-on.}
    \label{fig:2gpuExecDecomp}
\end{figure}

%% file: wrappers/4gpuExecDecomp.tex
\begin{figure}[t]
    \centering
    % Left subfigure (a)
    \begin{subfigure}{\columnwidth}
        \centering
        \includegraphics[width=\linewidth]{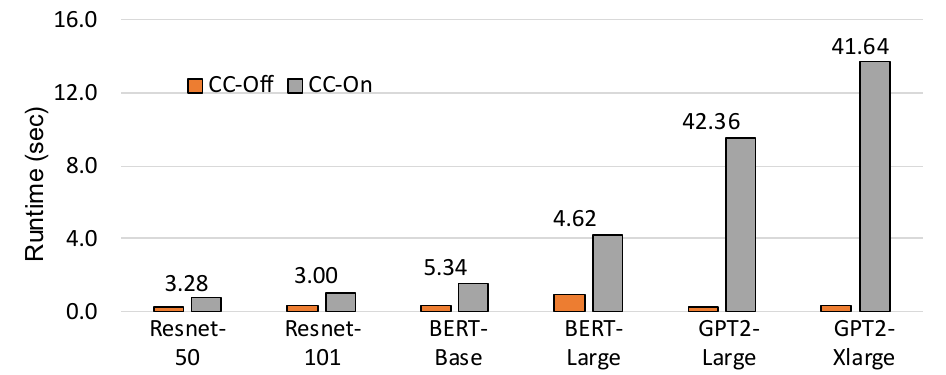}
        \caption{}
        \label{fig:4gpuexec}
    \end{subfigure}
    \hfill
    % Right subfigure (b)
    \begin{subfigure}{\columnwidth}
        \centering
        \includegraphics[width=\linewidth]{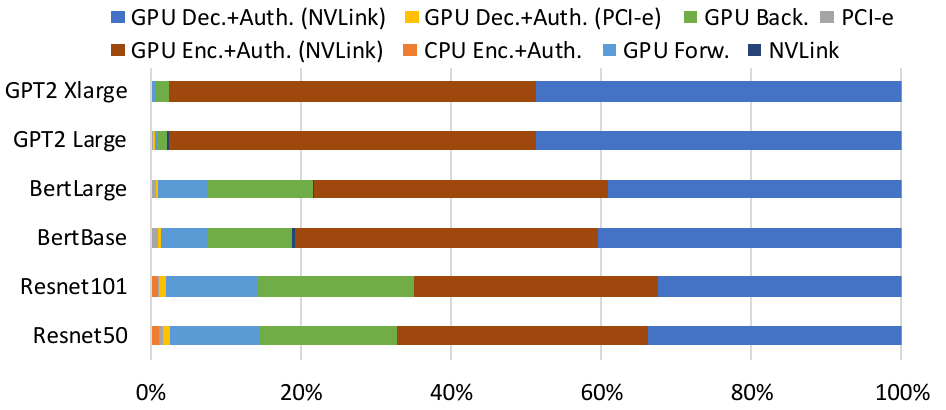}
        \caption{}
        \label{fig:4gpudecomp}
    \end{subfigure}
    \caption{DDP training on 4 GPU TEEs. (a) compares the per-training-iteration runtime (sec) with CC-off and CC-on. (b) demonstrates the runtime decomposition of 4 GPU training with CC-on.}
    \label{fig:4gpuExecDecomp}
\end{figure}

%% file: 7_relatedwork.tex
%This paper targets inter-GPU data movement in NVIDIA CC with multi-GPU settings, in which CVM with GPU TEEs manages secure communication. 
%\para{CPU-based TEEs} Intel SGX~\cite{sgx}, Intel TDX~\cite{tdx}, ARM TrustZone~\cite{trustzon}, and AMD SEV~\cite{amdcc}, provide additional hardware-based security measures but are limited in their parallelism. This limitation has led to offloading ML workloads to unsecured GPU~\cite{origami, tramer2019slalom, darkNight, kumar2020cryptflow}.

%\para{GPU-based TEEs} 

\bheading{GPU-based TEEs.}
%Graviton~\cite{graviton}, HIX~\cite{hix} provide GPU TEEs that extend TCB to GPU without modifying the GPU hardware. 
Based on NVIDIA CC-enabled H100~\cite{nvidiacc}, which is the first commercially available GPU TEE, \cite{mohan2024securing, zhu2024confidential} analyze the performance overhead of CPU-GPU communication in a CVM consisting of a single CC-enabled H100 GPU focusing on performance impact during LLM inference. %These works focus on the performance impact during LLM inference and report that, as model sizes increase, the performance degradation remains negligible compared to settings where CC is disabled.

Fastrack~\cite{wang2024fastrack} investigates CPU-GPU communication overhead in a single CC-enabled H100 GPU, focusing on training and inference of data-intensive ML applications such as ResNet, GraphSAGE, and Two-Tower Neural Networks~\cite{ttnn1,yi2019sampling}. The paper introduces optimization techniques: eliminating redundant CPU TEE encryption with a shared symmetric key between the client and GPU TEE, multi-chaining authentication to reduce GMAC generation overhead, and overlapping authentication with the AES decryption. These optimizations can also be applied to multi-GPU training in GPU TEEs.

Tan et al.~\cite{tan2024performance} characterizes the throughput reduction in cloud AI workloads such as TensorRT, PEFT, and vLLM due to secure CPU-GPU communication, and proposes to utilize multiple encryption workers to work in the background.PipeLLM~\cite{tan2024performance} reduces the latency overhead of LLM serving by pipelining encryption with GPU computation. The paper predicts which data requires encryption by analyzing the serving patterns of LLMs and launches an encryption request on the data speculatively.

Na et al.~\cite{na2024supporting} examines two key aspects of inter-GPU communication overhead in GPU TEEs: the cost of using the CPU-oriented shared-memory OTP generation scheme in multi-GPU environments and the additional bandwidth consumption caused by secure metadata, such as MACs and counters. %To address these challenges, the paper proposes optimizations such as dynamically managing AES pads for distributed settings and batching metadata. 
However, the evaluation is limited to small benchmarks, failing to reveal system bottlenecks. In contrast, our work comprehensively analyzes overall system performance when running large-scale ML models in distributed settings.

%\bheading{Trusted I/O and NVIDIA Protected PCIe.} 
\bheading{Trusted I/O.} AMD and Intel have also proposed a Trusted I/O (TIO)~\cite{intel:2023:tio,amd:2023:tio} standard to accelerate device-TEE performance by embedding dedicated encryption engines for PCIe streaming. This hardware-accelerated PCIe encryption offloads cryptographic work from the CPU, promising higher throughput and lower latency for GPU TEEs. Unfortunately, TIO and its companion protocol (TDISP) won’t be part of the PCIe specification until version 6.0, so today’s GPU-TEE solutions, even Azure’s commercial offerings, still run on PCIe 4.0/5.0 and cannot yet leverage the TIO performance boost. 

\bheading{NVIDIA Protected PCIe.} More recently, NVIDIA released Protected PCIe (PPCIe)~\cite{nvidiappcie}, an efficient multi-GPU training with minimal performance loss. However, PPCIe assumes that NVLink/NVSwitch connecting its GPUs are secure, removing encryption and authentication in GPU-to-GPU communications. Our paper is based on a stricter threat model, where we trust only the CPU and GPU packages assuming that only the data and code within the CPU and GPU TEEs are secure. Additionally, PPCIe is only enabled in a rigid constraint of a fixed 8-GPU configuration. This inflexibility hinders the adoption of secure multi-GPU training where it is more common setup in research clusters and clouds to use fewer GPUs as CC-available machines like H100 or B100 are very costly.

\begin{comment}
\section*{Acknowledgments}

The preferred spelling of the word ``acknowledgments'' in America is without 
an ``e'' after the ``g''. Avoid the stilted expression ``one of us (R. B. 
G.) thanks $\ldots$''. Instead, try ``R. B. G. thanks$\ldots$''. Put sponsor 
acknowledgments in the unnumbered footnote on the first page.
\end{comment}

%% file: 8_conclusion.tex
%In this work, we characterize the performance overhead introduced by GPU TEEs in DDP ML training. While GPU TEEs provide security features to protect data, models, and computation during training, they also reveal new performance challenges regarding gradient synchronization in distributed systems. Specifically, we quantify the TEE overhead in ring all-reduce operations used in DDP.
In this work, we characterize the performance overhead introduced by GPU TEEs in DDP ML training, specifically, over the ring all-reduce operation used to synchronize gradients between multiple GPUs.
We recognize the three major bottlenecks that increase TEE overheads. Firstly, a single ring all-reduce operation is implemented with multiple sub-operations that incur encryption and authentication. Secondly, as the number of GPUs increases, the steps of sub-operations to complete a single all-reduce also increase linearly, resulting in more TEE overheads in a single training iteration. Thirdly, the model size determines the number of asynchronous all-reduce operations launched within the backward propagation of training. These factors collectively contribute to substantial runtime slowdowns, with a maximum observed runtime increase of 41.64x in GPT2-Xlarge and an average of 8.68x for DDP training with four GPU TEEs compared to systems without any security measures.
Lastly, we propose batching these asynchronous all-reduces by modifying the DDP hyperparameter bucket\_cap\_mb to mitigate TEE overheads. While we significantly reduce the TEE overheads, secure four-GPU DDP with optimal bucket\_cap\_mb still suffers from 3.03x increased runtime compared to the non-TEE setting. This work emphasizes the critical trade-off between security and performance in distributed ML systems, emphasizing the need for optimized training hyperparameters to mitigate GPU TEE overheads. 